\documentclass{appolb}
\usepackage{epsfig}

\begin{document}
\title{High $p_T$ jets in DIS and $\gamma p$ at HERA%
\thanks{Presented at ISMD2007, Berkeley CA}%
}
\author{N. Coppola, on behalf of the H1 and ZEUS Collaborations
\address{Deutsches Elektron-Synchrotron DESY, Notkestrasse 85, 22607
  Hamburg, Germany}%
}
\maketitle
\begin{abstract}
  Recent results on di-jet, inclusive jets, production at high $p_T$
  in Deep Inelastic Scattering DIS and photoproduction $\gamma p$
  regimes using both the H1 and ZEUS detectors at HERA are presented.
  Also studies of integrated jet shape and jet radius dependencies in
  DIS performed by the ZEUS Collaborations are discussed.  All the
  measurements are found to be well described by calculations at the
  next-to-leading order in perturbative QCD.  A combined
  determination of the value of the strong coupling constant
  $\alpha_s(M_Z)$ from the H1 and ZEUS Collaborations using
  inclusive-jet cross-section measurements in neutral current DIS at
  high $Q^2$ is shown.
\end{abstract}
\PACS{PACS numbers come here}
  
\section{Introduction}
Measurements of jets cross-sections in high-$Q^2$ deep-inelastic
scattering (DIS) as well as in photoproduction $\gamma p$ interactions
have traditionally been used to test the concepts of perturbative QCD
(power expansion, factorisation, PDF universality). In addition, jet
measurements in DIS allow precise determinations of the strong
coupling $\alpha_S$ and are a valuable input to global fits of the
PDFs (see for example~\cite{pdfs}). In this article new results
published by the H1 and ZEUS collaborations will be presented.

\section{Di-jet cross-section}
\begin{figure}[t]
  \vspace{-0.5cm}
  \begin{minipage}[h]{13.1cm}
  \centering
  \hspace{-0.6cm}\psfig{figure=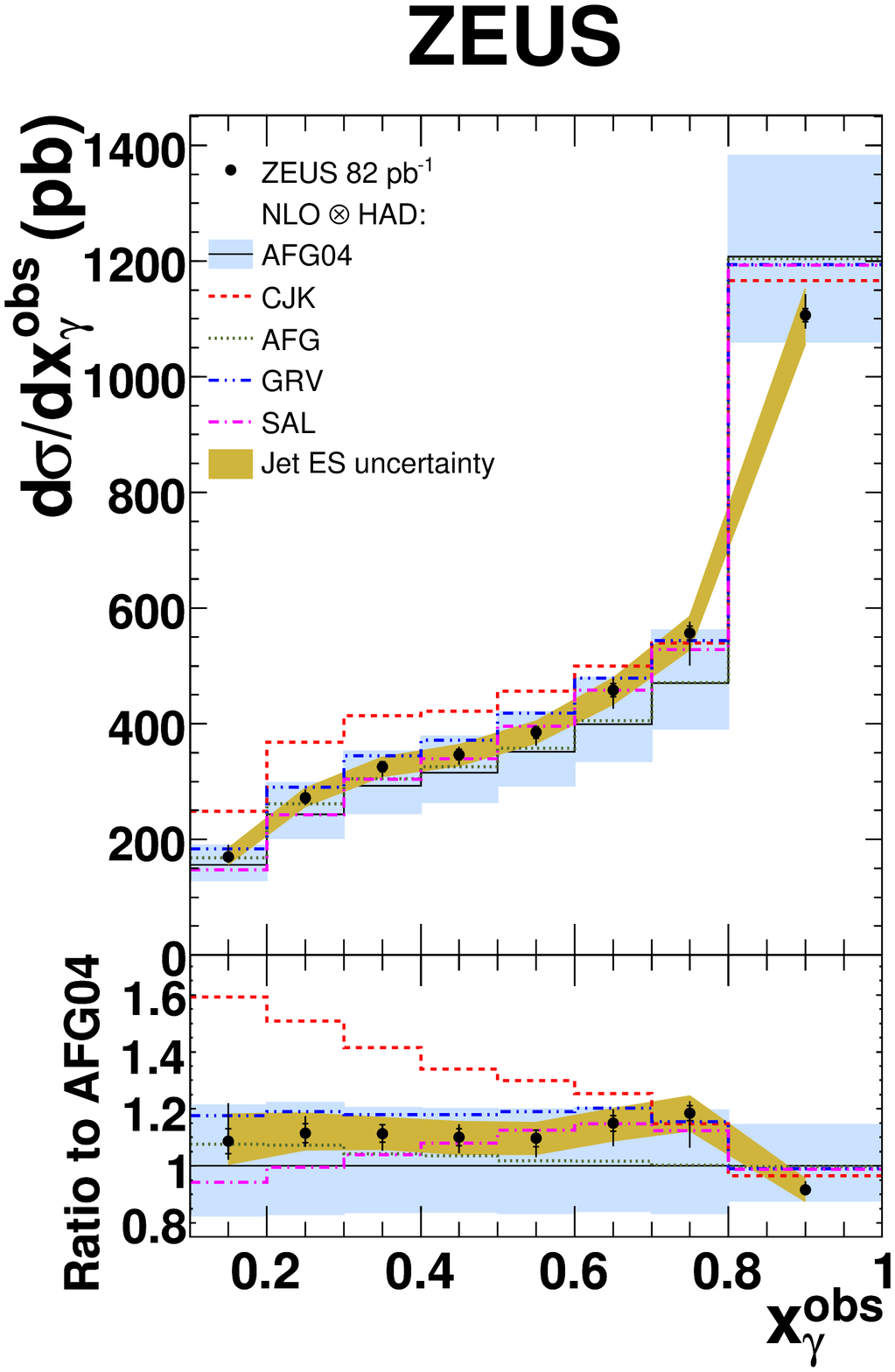,height=5.9cm}
  \vspace{-6.4cm}
  \hspace{0.1cm}
  \psfig{figure=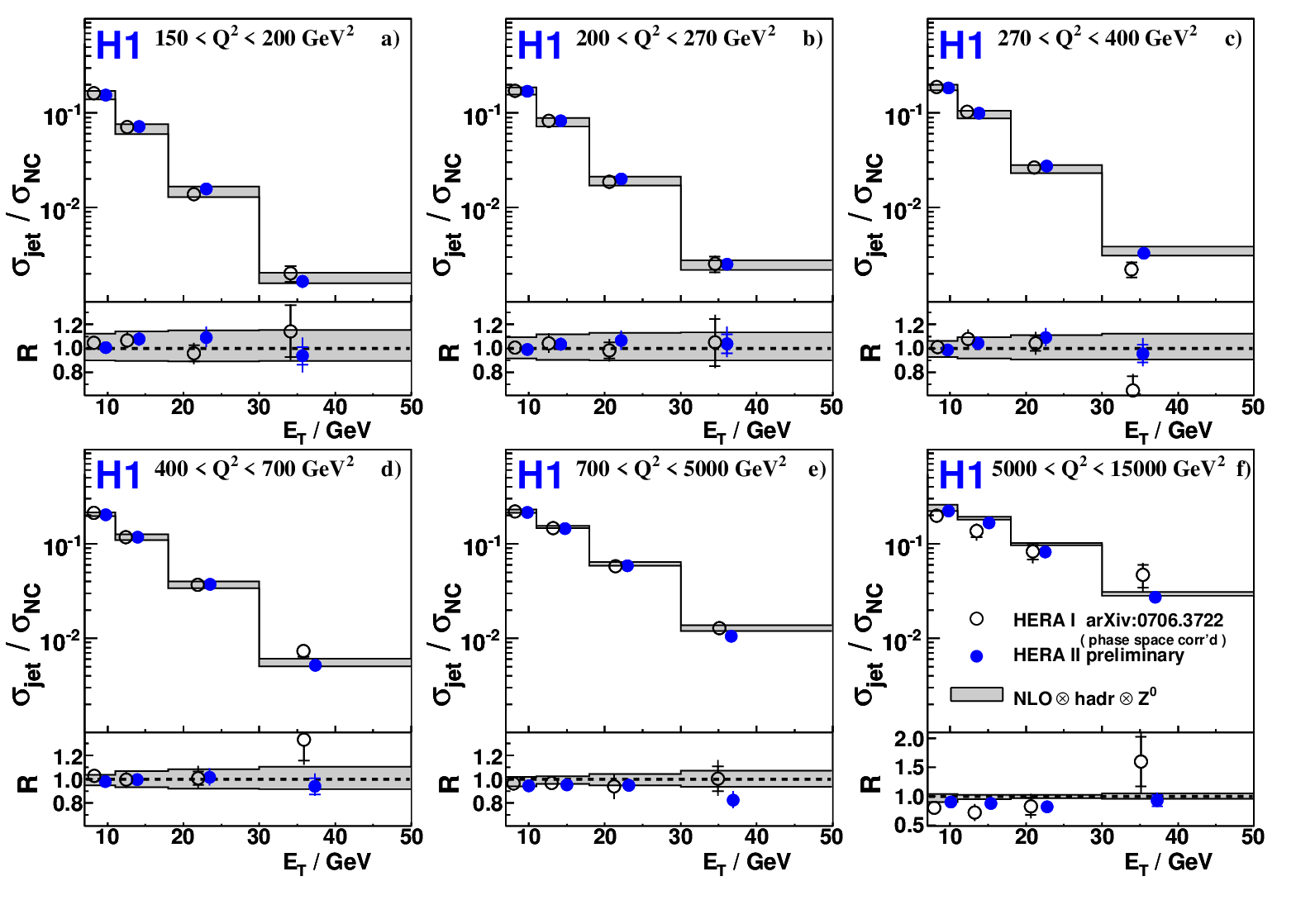,height=5.3cm}
  \vspace{6.4cm}
  \end{minipage}
  \caption{(left) Di-jets cross-section $d\sigma/d
    {x_\gamma}^{\mbox{\scriptsize obs}}$ by ZEUS. Data are compared
    with the results of NLO by Frixione Ridolfi obtained using
    numerous photon PDFs.  (right) Inclusive-jets double-differential
    cross-section $d^2\sigma/dE^{\mbox{\scriptsize jet}}_{J,B} dQ^2$
    by H1. Data are compared with the results of NLOJET++, corrected
    for hadronisation and $Z^0$ boson exchange.}
  \label{fig:incljets}
  \label{fig:dijet}
\end{figure}

Measurements of cross-sections for high-$E_T$ di-jet in
photoproduction, when a quasi-real photon, emitted from the incoming
electron, collides with a parton from the incoming proton, are
presented. The data samples were collected with the ZEUS detector
and correspond to an integrated luminosity of ${\cal
  L}=81.8~\mbox{pb}^{-1}$~\cite{dijet-zeus}. The measured
cross-sections show a sensitivity to the parton distributions in the
photon (see Fig.~\ref{fig:dijet} (left)) and in the proton and the QCD
effects beyond next-to-leading order. The data are therefore
well-suited to further constrain the proton and photon distribution
functions when used in?? global QCD fits (the same conclusions can be
derived by looking at H1 data as
in~\cite{slides}~and~\cite{h1phpoptimised}).

\section{Inclusive-jet cross-section}
Inclusive-jet cross-section based on the total HERA-II data
set equivalent to ${\cal L}\simeq 320~\mbox{pb}^{-1}$ recorded by the
H1 detector have been reported. The inclusive-jet cross-section
normalized to the neutral current deep inelastic scattering (NC DIS)
cross-section is measured as a function of the hard scale $Q^2$ and
$E_{T,B}$, the transverse jet energy in the Breit frame~\cite{breit},
in the phase space defined by $150<Q^2< 15000~\mbox{GeV}^2$,
$0.2<y<0.7$, where $y$ quantifies the inelasticity of the interaction,
and for $7<E_{T,B}<50~\mbox{GeV}$. The measurements are found to be
well described by calculations at the NLO in perturbative QCD, pQCD.
The present preliminary results are compared to recently published
results based on 1999-2000 data recorded by H1 during the HERA-I
period with ${\cal L}=65.4~\mbox{pb}^{-1}$. As can be seen in
Fig.~\ref{fig:incljets} (right) the use of a much larger dataset
significantly improves the statistical precision and reduces the total
uncertainty of the measurements.

\section{Jet-radius dependence and integrated jet shape}
At sufficiently high $E_T^{\mbox{\scriptsize jets}}$, where
fragmentation effects become negligible, the jet substructure is
expected to be calculable by pQCD. Moreover, measurements of jet
substructure allow investigations of the differences between quark and
gluon initiated jets and the dynamics of the different partonic final
states, as well as determinations of $\alpha_S$.

Studies of the jet-radius dependency performed with an inclusive-jet
sample in DIS with $Q^2>125~\mbox{GeV}^2$ have shown that there is a
linear proportionality of the radius $R$ on the differential cross
section as function of both $Q$ and $E_{T,B}^{\mbox{\scriptsize jet}}$
for $0.5\le R\le 1.$~\cite{jet-radius}. Similarly, an analysis has
been performed using DIS events, in the same phase space, collected
during the HERA-II period with the ZEUS detector with ${\cal
  L}=368~\mbox{pb}^{-1}$. The data were divided into two samples with
one and two jet(s) of transverse energy $E_T>14~\mbox{GeV}$
respectively. As differentiating quantity, the integrated jet shape has
been used, defined as
$$
\left<\Psi(r)\right> = \frac{1}{N_{\mbox{\scriptsize jets}}}
\sum_{\mbox{\scriptsize jets}} \frac{E_T(r)}{E_T^{\mbox{\scriptsize jets}}}
$$
that is the average fraction of the jet's transverse energy that
lies inside a circle in the $\eta-\phi$ plane of radius $r$ concentric
with the jet axis. The results are compatible with the predictions of
pQCD that foresee that gluon jets are broader than quark jets
$\Psi_{\mbox{\scriptsize quarks}}(r)>\Psi_{\mbox{\scriptsize
    gluons}}(r)$ as can be seen in Fig.~\ref{fig:intjetshapes} (left).

\section{Combined $\alpha_S$ determination}
The strong coupling constant, $\alpha_S$, is one of the fundamental
parameters of QCD. However, its value is not predicted by theory and
must be determined experimentally. The success of perturbative QCD is
strengthened by precise and consistent determinations of the strong
coupling constant from many diverse phenomena~\cite{pdg}. New
determinations of the $\alpha_S$ have been recently published by the
H1~\cite{h1alphas} and ZEUS~\cite{zeusalphas} Collaborations. These
determinations have been performed from the measurements of
inclusive-jet cross-sections in NC DIS at high-$Q^2$ and have been
further used to perform a combined analysis and to obtain a single
more precise $\alpha_S$ value~\cite{combined-alpha}.

\begin{figure}[tb]
  \vspace{-0.5cm}
  \begin{minipage}[h]{13.0cm}
  \centering
  \hspace{-0.6cm}
  \psfig{figure=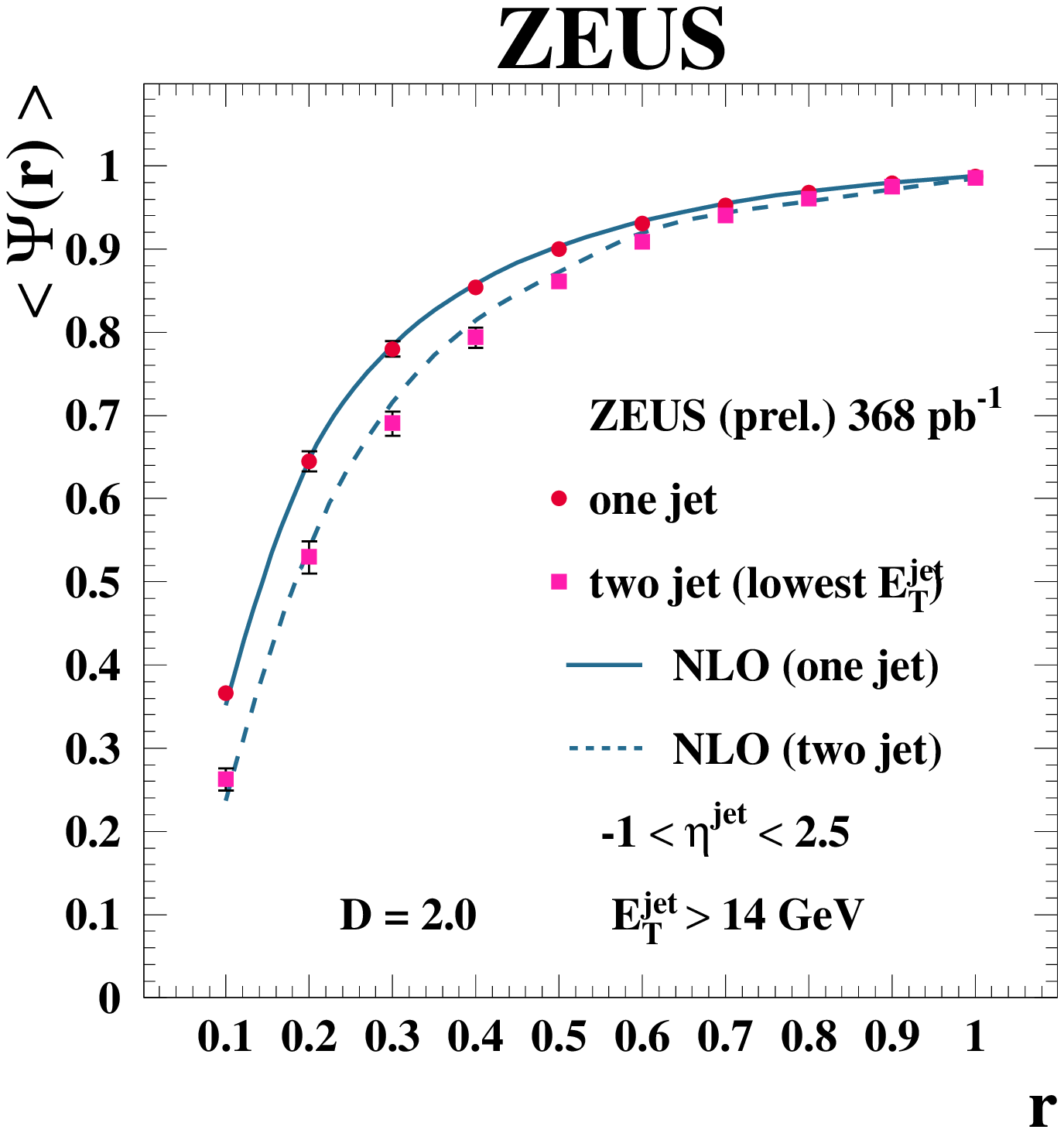,height=5.5cm}
  \hspace{-0.4cm}
  \psfig{figure=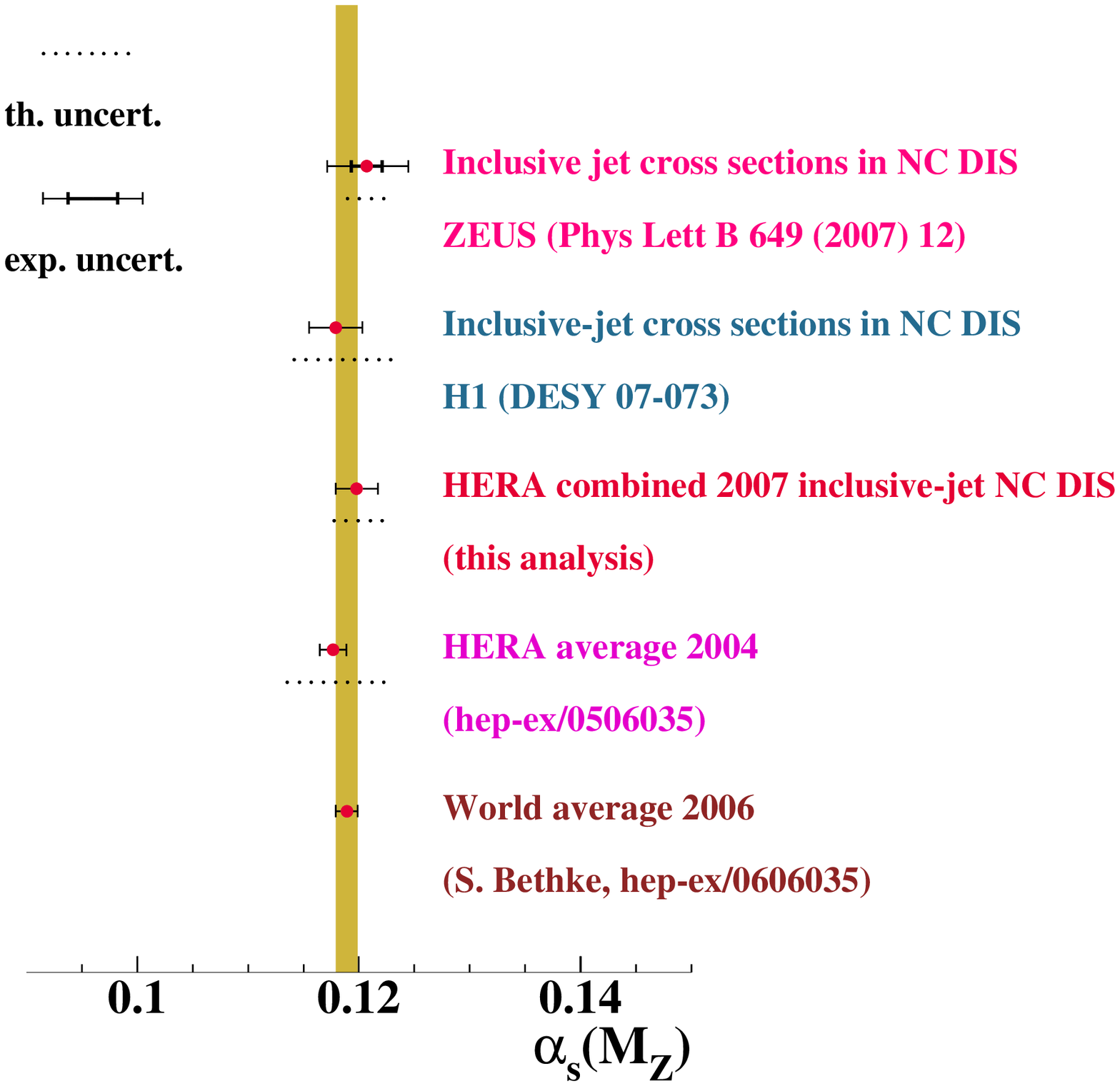,height=5.5cm}
  \end{minipage}
  \caption{(left) Integrated jet shape, points are ZEUS measurements,
    curves are NLO calculation with DISENT (1 jet events) and NLOJET++
    (2 jets events).  (right) $\alpha_S(M_Z)$ determinations from the
    H1 and ZEUS collaborations. The HERA combined 2007 $\alpha_S(M_Z)$
    and the HERA-2004 and the world-average are also shown.}
  \label{fig:alphastrong}
  \label{fig:intjetshapes}
\end{figure}

A simultaneous fit to the actual cross-section measurements, instead
of combining $\alpha_S$ values as it was done for the HERA-2004
average, was performed. The simultaneous fit was done to 24 H1 data
points in the range $150<Q^2< 15000~\mbox{GeV}^2$ (see
Fig.~\ref{fig:incljets} (right)) and 6 ZEUS data points in the range
$125<Q^2< 10000~\mbox{GeV}^2$. The NLO calculations used were based on
the MRST2001 PDF sets. The renormalisation and factorisation scales
were set to $E_{T,B}^{\mbox{jet}}$ and $Q$ respectively. The
experimental uncertainty on the combined $\alpha_S$ value amounts to
0.0019 and was obtained with the Hessian method, which fits the
sources of systematic uncertainties as the energy scale, luminosity,
model dependence, etc. The sources of systematic uncertainty were
treated as correlated for measurements within one experiment, but
uncorrelated between the two experiments. The theoretical uncertainty
coming from terms beyond NLO was estimated using the method of Jones
{\em et al.}~\cite{jones}, and gives the largest contribution. The
other sources of theoretical uncertainty considered were: PDF,
factorization scale and hadronisation. Therefore, the HERA combined
2007 $\alpha_S(M_Z)$ value is
$$
\alpha_S(M_Z)=0.1198\pm 0.0019 \mbox{(exp.)} \pm 0.0026 \mbox{(th.) 
(HERA combined 2007).} 
$$
This combined value is shown in Fig.~\ref{fig:alphastrong} (right)
together with the individual values obtained by both collaborations,
the HERA-2004 ($0.1186\pm 0.0011 \mbox{(exp.)} \pm 0.0050
\mbox{(th.)}$) and the world-2006 ($0.1189\pm 0.0010$) averages. The
measurements are consistent with each other and with the world
average. The HERA 2007 combined $\alpha_S(M_Z)$ has a much smaller
theoretical uncertainty due to the combination of the measurements in
which the theoretical uncertainties are well under control, at the
expense of a slight increase in the experimental uncertainty. This
value of $\alpha_S(M_Z)$ is very competitive with the most recent
result from LEP~\cite{lep}.

\end{document}